\documentclass[preprintnumbers,nofootinbib,noshowpacs,eqsecnum,twocolumn,superscriptaddress,prd]{revtex4}

\usepackage{graphicx}
\usepackage{amsmath,amsthm,amssymb}
\usepackage{mathrsfs,bbm}
\usepackage{color}

\definecolor{lgray}{gray}{0.9} 		
\graphicspath{ {figures/} }

\makeatletter       

\renewcommand{\p@subsection}{}

\makeatother

\newcommand*{\eweakgroup}{\mbox{$SU(2)_L \times U(1)_Y$} }
\newcommand*{\emgroup}{\mbox{$U(1)_{em}$} }

\newcommand*{\unitmatrix}{\mathbbm{1}}

\newcommand*{\twomat}[1]{\underline{#1}}             
\newcommand*{\tvec}[1]{\boldsymbol{#1}}              
\newcommand*{\trans}{\mathrm{T}}                     
\newcommand*{\by}{\!\times\!}                        
\newcommand*{\im}[1]{\text{Im} {#1}}                        
\newcommand*{\re}[1]{\text{Re} {#1}}

\DeclareMathOperator{\trace}{tr}

\DeclareMathOperator{\diag}{diag}		

\begin{document}

\preprint{ADP-15-11/T913}
\title{Stability and symmetry breaking in a three-Higgs-doublet model with
lepton family symmetry $O(2) \otimes \mathbbm{Z}_2$}

\author{M. Maniatis}
    \email[E-mail: ]{MManiatis@ubiobio.cl}
\affiliation{Departamento de Ciencias B\'a{}sicas, 
Universidad del B\'i{}o B\'i{}o, Casilla 447, Chill\'a{}n, Chile.}
\author{D. Mehta}
\email[E-mail: ]{dmehta@nd.edu}
\affiliation{Department of Applied and Computational Mathematics and Statistics, University of Notre Dame, Notre Dame, IN 46556, USA}
\affiliation{Centre for the Subatomic Structure of Matter, Department of Physics, School of Physical Sciences,
University of Adelaide, Adelaide, South Australia 5005, Australia.}

\author{Carlos M. Reyes}
\email[E-mail: ]{creyes@ubiobio.cl}
\affiliation{Departamento de Ciencias B\'a{}sicas, 
Universidad del B\'i{}o B\'i{}o, Casilla 447, Chill\'a{}n, Chile.}
%

\begin{abstract}
Motivated by the neutrino data, an extension of the Standard Model
with three Higgs-boson doublets has been proposed.
Imposing an $O(2) \otimes \mathbbm{Z}_2$ family symmetry, 
a neutrino mixing matrix with 
$\theta_{23}=\pi/4$ and $\theta_{13}=0$ appears in a natural way.
Even though these values for the mixing matrix do not follow the
recent experimental constraints, they are nevertheless a good approximation.
We study the Higgs potential of this model in detail. 
We apply recent methods which allow for the study of any three-Higgs-boson doublet
model. It turns out that for a variety of parameters the potential
is stable, has the correct electroweak symmetry breaking, and gives the 
correct vacuum expectation value.
\end{abstract}


\maketitle

  
\section{The $O(2) \otimes \mathbbm{Z}_2$ model}
\label{model}


The experimental neutrino mixing data show that the neutrino mixing
is very different from the quark mixing. In the usual 
parametrization of the neutrino mixing matrix (see for instance~\cite{Agashe:2014kda}), experimental 
data suggest that the angle $\theta_{13}$ is small (but nonzero),
and $\theta_{23}$ is close to $\pi/4$~\cite{GonzalezGarcia:2012sz}.

There is lot of effort spent on finding an organizing principle for the
flavor puzzle. A general approach is to study finite subgroups of
SU(3) which have an irreducible triplet representation; see for instance \cite{King:2013eh}
and references therein.
Examples of subgroups with triplet representations are $S_4$,
the group of
permutations of 4 elements, with two singlet, one doublet, and two triplet 
representations (see for instance \cite{Ma:2005pd});
or $A_4$, 
the group of even permutations of 4 elements which has also
a triplet representation besides singlet representations
(see for instance~\cite{Ma:2009wi,Felipe:2013vwa}); 
or $\Delta(27)$, with
27 elements and two triplet representations
(see for instance \cite{Ma:2013xqa}). 

In contrast, here we want to 
study in detail the Higgs potential of a three Higgs-boson-doublet model which 
imposes a $O(2) \times \mathbbm{Z}_2$ symmetry -- without any irreducible triplet representation. 
Let us closely follow the motivation of~\cite{Grimus:2008dr}. 
The starting point is a neutrino mass matrix which is
symmetric in generations two and three,
\begin{equation} \label{massmatrix}
M_\nu =
\begin{pmatrix}
x & y & y\\
y & z & w\\
y & w & z
\end{pmatrix}.
\end{equation}
This mass matrix may be diagonalized as usual, that is, 
$U^\dagger M_\nu U = \diag(m_{\nu_1}^2, m_{\nu_2}^2, m_{\nu_3}^2)$,
where $U$ is the neutrino mixing matrix and $m_{\nu_1}$, $m_{\nu_2}$, $m_{\nu_3}$ the
neutrino masses.
Expressing the mixing matrix $U$ in terms of the 
usual parametrization~\cite{Agashe:2014kda}, we get in particular $\theta_{13}=0$ and
$\theta_{23}= \pi/4$. 
Even though the experimental results are not
in exact agreement with these values, in particular $\theta_{13}$ 
is nonzero, they at least appear to be
approximately fulfilled. 
We note that deviations from these approximate values for 
the mixing angles could arise by soft-symmetry-breaking
terms or beyond-leading-order
effects; see for instance the study in \cite{Altarelli:2010gt}.

The mass matrix~\eqref{massmatrix} may be generated by the introduction of
three Higgs-boson doublets~$\varphi_i$, $i=1,2,3$, and a symmetry $O(2) \times \mathbbm{Z}_2 \cong 
 \mathbbm{Z}_2' \times \mathbbm{Z}_2 \times U(1)$,
where all elementary particles are assigned to an appropriate 
transformation behavior on this symmetry. 
The reflection symmetry~$\mathbbm{Z}_2'$ is 
responsible for the $\mu$--$\tau$ symmetry of $\eqref{massmatrix}$:
\begin{alignat}{2} \label{s}
&\mathbbm{Z}_2': \quad
&&
D_{\mu L} \leftrightarrow D_{\tau L}, \;
\mu_R  \leftrightarrow \tau_R, \;
\nu_{\mu R} \leftrightarrow \nu_{\tau R}, \;
\phi_1 \leftrightarrow \phi_2\\
\intertext{Here $D_{\mu L}$ and $D_{\tau L}$ denote the left-handed $SU(2)$ lepton doublets,
$\nu_{\mu R}$ and $\nu_{\tau R}$ the right-handed neutrinos, and
all remaining fields transform trivially under the $\mathbbm{Z}_2'$ symmetry.
The $\mathbbm{Z}_2$ symmetry is given by a sign change,
}
\label{Z2}
&\mathbbm{Z}_2: \nonumber
&&\nu_{e R}  \rightarrow -\nu_{e R}, \;
\nu_{\mu R}  \rightarrow -\nu_{\mu R}, \;
\nu_{\tau R}  \rightarrow -\nu_{\tau R}, \\
&
&&
e_R \rightarrow -e_R, \;
\varphi_3 \rightarrow -\varphi_3, 
\intertext{with $\nu_{e R}$, $\nu_{\mu R}$, $\nu_{\tau R}$ the right-handed neutrinos, $e_R$ the right-handed
electron,
and all other fields unchanged under this $\mathbbm{Z}_2$ symmetry.
Eventually  the assignment with respect to the phase symmetry $U(1)$ is
}
\label{U1}
&U(1):
&&
\begin{array}{c|cccc}
X \longrightarrow e^{i \theta} X & 
D_{\mu L}, \tau_R, \nu_{\mu R} &\, D_{\tau L}, \mu_R, \nu_{\tau R} &\, \varphi_1 &\, \varphi_2\\
\hline
\theta & 1 &  -1 & 2 &  -2
\end{array}
\end{alignat}
with~$X$ one of the fields on the right-hand side of the table transforming as
$X \longrightarrow e^{i \theta} X$ with the corresponding phase~$\theta$ given explicitly 
in the table. All other fields transform trivially.

By virtue of these symmetries -- besides the electroweak \eweakgroup symmetry -- there appear in particular the invariant  
Yukawa couplings 
\begin{equation} \label{Yukawa}
{\mathscr L}_Y = 
- y_4 \left( \bar{D}_{\mu L} \varphi_1 \mu_R + \bar{D}_{\tau L} \varphi_2 \tau_R  \right) + h.c.
\end{equation}

The most general potential for the three Higgs-boson doublets
$\varphi_1$, $\varphi_2$, $\varphi_3$ reads
\begin{multline} \label{VO2}
V_{O(2)\times \mathbbm{Z}_2} = \mu_0 \varphi_3^\dagger \varphi_3 
+\mu_{12} \left( \varphi_1^\dagger \varphi_1 + \varphi_2^\dagger \varphi_2 \right)\\
+\mu_m \left( \varphi_1^\dagger \varphi_2 + \varphi_2^\dagger \varphi_1 \right)
+a_1 (\varphi_3^\dagger \varphi_3)^2\\
+a_2 \varphi_3^\dagger \varphi_3 
 \left( \varphi_1^\dagger \varphi_1 + \varphi_2^\dagger \varphi_2 \right)\\ 
+a_3 \left( \varphi_3^\dagger \varphi_1 \cdot \varphi_1^\dagger \varphi_3 +
	\varphi_3^\dagger \varphi_2 \cdot \varphi_2^\dagger \varphi_3  \right) \\
+a_4 \varphi_3^\dagger \varphi_1 \cdot \varphi_3^\dagger \varphi_2
+a_4^* \varphi_1^\dagger \varphi_3 \cdot \varphi_2^\dagger \varphi_3 \\
+a_5 \left( (\varphi_1^\dagger \varphi_1)^2 + (\varphi_2^\dagger \varphi_2)^2 \right)\\
+a_6 \varphi_1^\dagger \varphi_1 \cdot \varphi_2^\dagger \varphi_2
+a_7 \varphi_1^\dagger \varphi_2 \cdot \varphi_2^\dagger \varphi_1,
\end{multline}
where the term 
$\mu_m (\varphi_1^\dagger \varphi_2 + \varphi_2^\dagger \varphi_1)$
breaks the $U(1)$ symmetry \eqref{U1} explicitly but softly, since this is
a quadratic term.
For a nonvanishing parameter $\mu_m$ in this way additional Goldstone bosons are avoided,
which otherwise would appear by spontaneous symmetry breaking of the
$U(1)$ symmetry. This potential has nine real parameters and one complex parameter~$a_4$, corresponding
to eleven real parameters in total.

Now we want to discuss stability, stationarity, and electroweak symmetry breaking of this model. Of course
only a model with a stable potential, having the correct-electroweak-symmetry-breaking behavior 
and the correct vacuum expectation values is physically acceptable.
These obvious constraints restrict the parameter space of the potential. 
Here, we focus on the Higgs potential and not on any further experimental limits.
For instance, the expressions for the 
{\em oblique parameters} $S$, $T$, $U$ are available for any nHDM~\cite{Grimus:2007if}
and may be compared to the electroweak precision data.

Since the $O(2) \otimes \mathbbm{Z}_2$ model is a
3HDM we encounter in this model four charged Higgs bosons and in total five neutral Higgs bosons.
We expect a changed phenomenology of this model compared to the Standard Model.
Of course, the detection of any further Higgs boson would be a clear signal for a model beyond the Standard Model. 
Depending on the Yukawa coupling strength in (1.5) we have for instance the signature of 
the production of a charged Higgs boson with subsequent
decay into a muon and a muon-neutrino. The potential itself is in principle
detectable via its trilinear and quartic Higgs self-couplings. 
We leave this investigation for future work and focus here on the study of the Higgs potential
with respect to stability, electroweak symmetry breaking and the global minimum.

Even though the potential~\eqref{VO2} appears to be rather involved we will see
that it is indeed accessible in the bilinear approach~\cite{Nagel:2004sw,Maniatis:2006fs,Nishi:2006tg}.
In this approach gauge degrees of freedom are avoided systematically.
Moreover, the corresponding equations for stability and stationarity simplify in
particular, the degree of systems of equations is lowered.
Recently, the bilinear approach for the study of stability, stationarity,
and electroweak symmetry breaking has been extended to the study of any
3HDM~\cite{Maniatis:2014oza}, which we now briefly review.

The scalar products of the type $\varphi_i^\dagger \varphi_j$, $i,j \in \{1,2,3\}$, in the potential~\eqref{VO2} may be 
arranged in a $3\by 3$ matrix
\begin{gather}
\label{eq-kmat}
\twomat{K} =
\begin{pmatrix}
  \varphi_1^{\dagger}\varphi_1 & \varphi_2^{\dagger}\varphi_1 & \varphi_3^{\dagger}\varphi_1 \\
  \varphi_1^{\dagger}\varphi_2 & \varphi_2^{\dagger}\varphi_2 & \varphi_3^{\dagger}\varphi_2 \\
  \varphi_1^{\dagger}\varphi_3 & \varphi_2^{\dagger}\varphi_3 & \varphi_3^{\dagger}\varphi_3   
\end{pmatrix}.
\end{gather}
By the introduction of the bilinears,
\begin{equation} \label{2.6}
K_\alpha = K_\alpha^* = \trace (\twomat{K} \lambda_\alpha), \qquad \alpha=0,...,8
\end{equation}
with $\lambda_\alpha$ the $3\by 3$~Gell-Mann matrices,
the following replacements can be made in the potential:
\begin{alignat}{2} 
\nonumber
&\varphi_1^\dagger \varphi_1 = \frac{K_0}{\sqrt{6}} +\frac{K_3}{2} +\frac{K_8}{2\sqrt{3}},\qquad
&&\varphi_1^\dagger \varphi_2 = \frac{1}{2}\left( K_1 + i K_2 \right),\\
\label{phiK}
&\varphi_1^\dagger \varphi_3 = \frac{1}{2}\left( K_4 + i K_5 \right), \qquad
&&\varphi_2^\dagger \varphi_2 = \frac{K_0}{\sqrt{6}} -\frac{K_3}{2} +\frac{K_8}{2\sqrt{3}},\\
\nonumber
&\varphi_2^\dagger \varphi_3 = \frac{1}{2}\left( K_6 + i K_7 \right), \qquad
&&\varphi_3^\dagger \varphi_3 = \frac{K_0}{\sqrt{6}} - \frac{K_8}{\sqrt{3}}.
\end{alignat}
Comparing the potential, written in terms of bilinears 
\begin{equation}
\label{eq-vdef}
V = \xi_\alpha K_\alpha + K_\alpha \tilde{E}_{\alpha \beta} K_\beta, \quad \alpha, \beta = (0,\ldots, 8),
\end{equation}
with the general form of the potential,
we find the new parameters
\begin{widetext}
\begin{equation} \label{paraK}
\begin{split}
&\left(\xi_\alpha \right) = 
\left(\frac{\mu_0 + 2 \mu_{12}}{\sqrt{6}}, \mu_m, 0, 0, 0, 0, 0, 0, \frac{\mu_{12}-\mu_0}{\sqrt{3}}\right)^\trans, \\
&\left( \tilde{E}_{\alpha \beta} \right) = \frac{1}{4} 
\cdot
\begin{pmatrix}
\frac{2}{3} (a_1 + 2 a_2 + 2 a_5 + a_6) &
0 & 0& 0& 0& 0& 0& 0& \frac{\sqrt{8}}{3}(-a_1-\frac{a_2}{2}+a_5+\frac{a_6}{2})\\
0 & a_7 & 0 & 0 & 0 & 0 & 0 & 0 & 0\\
0 & 0 & a_7 & 0 & 0 & 0 & 0 & 0 & 0\\
0 & 0 & 0 & 2 a_5-a_6 & 0 & 0 & 0 & 0 & 0\\
0 & 0 & 0 & 0 & a_3 & 0 & \re a_4 & \im a_4 & 0\\
0 & 0 & 0 & 0 & 0 & a_3 & \im a_4 & -\re a_4 & 0\\
0 & 0 & 0 & 0 & \re a_4 & \im a_4 & a_3 & 0 & 0\\
0 & 0 & 0 & 0 & \im a_4 & -\re a_4 & 0 & a_3 & 0\\
\frac{\sqrt{8}}{3}(-a_1-\frac{a_2}{2}+a_5+\frac{a_6}{2}) & 0 & 0 & 0 & 0 & 0 & 0 & 0 & \frac{4}{3} a_1 - \frac{4}{3} a_2+ \frac{2}{3} a_5+ \frac{1}{3} a_6
\end{pmatrix}.
\end{split}
\end{equation}
\end{widetext}
Obviously, all parameters are real in terms of bilinears. 
We note that there is a one-to-one correspondence between the Higgs-boson doublets
and the bilinear matrix $\twomat{K}=1/2 K_\alpha \lambda_\alpha$
with rank smaller or equal to two - except for irrelevant gauge degrees of freedom; 
see \cite{Maniatis:2006fs}.

Supposing the potential is bounded from below, 
at the global minimum, or the degenerate minima, the gradient of the potential has to vanish.
The corresponding equations may be used to fix some of the parameters.
In order to obtain these equations we start with the parametrization of the three Higgs-boson doublets
with the same hypercharge $y=+1/2$, in a particular gauge,
\begin{equation}\label{expandV}
\begin{split}
&\varphi_{1/2}(x) = 
\begin{pmatrix}
\phi^+_{1/2}(x)\\
\frac{1}{\sqrt{2}} ( v_{1/2} + H^0_{1/2}(x) + i A^0_{1/2}(x))
\end{pmatrix},\\
& \varphi_3(x) = 
\begin{pmatrix}
0\\
\frac{1}{\sqrt{2}} ( v_3 + h^0(x) )
\end{pmatrix}.
\end{split}
\end{equation}
The derivatives of the potential~\eqref{VO2}, inserting \eqref{expandV} with respect to
the  fields
at the vacuum, that is, for vanishing fields give the nontrivial conditions
\begin{equation}\label{tadpole}
\begin{split}
&\mu_0 = - \frac{1}{2}(a_2+a_3)(v_1^2+v_2^2) -a_1 v_3^2 - \re(a_4) v_1 v_2,\\
&\mu_{12} =  - \frac{1}{2} (a_2+a_3) v_3^2 - a_5 (v_1^2 + v_2^2),\\
&\mu_m = \frac{1}{2}\left((2 a_5 - a_6 - a_7) v_1 v_2 - \re(a_4) v_3^2\right),\\
&\im(a_4) \cdot v_2  v_3^2 =0,\\
&\im(a_4) \cdot v_1  v_3^2 =0.
\end{split}
\end{equation}
For nonvanishing vacuum expectation values, the last two
equations immediately dictate that~$a_4$ has to be real.
Eventually, by means of the equations~\eqref{tadpole} the
quadratic parameters $\mu_0$, $\mu_{12}$, $\mu_m$ may be expressed by the
quartic parameters and the three vacuum expectation values $v_1$, $v_2$, $v_3$.
Further, the vacuum expectation values are restricted with view on the 
Yukawa couplings~\eqref{Yukawa}, that is, the ratio of the vacuum expectation values $v_1$ and
$v_2$ has to be  $v_1/v_2 = m_\mu/m_\tau$
at tree-level accuracy. In addition, the vacuum expectation value 
\begin{equation} \label{vev}
v_0 = \sqrt{v_1^2+v_2^2+v_3^2} \approx 246~\text{GeV}
\end{equation}
is given by the electroweak 
precision data. Therefore, all quadratic parameters follow from the quartic parameters
and one free vacuum expectation value, say~$v_3$.
Therefore, it appears reasonable to start with the following set of parameters,
\begin{equation} \label{newparameters}
\begin{split}
&a_1, a_2, a_3, \re(a_4), a_5, a_6, a_7, v_3,\\
&v_0 \approx 246~\text{GeV}, v_1/v_2 = m_\mu/m_\tau.
\end{split}
\end{equation}
Note that the tadpole conditions~\eqref{tadpole} only ensure that 
there is at least one stationary solution.
By no means does this guarantee that
the corresponding potential is stable and has 
a global minimum with the correct partially broken electroweak
symmetry. 

  
\section{Stability and electroweak symmetry breaking in the $O(2) \otimes \mathbbm{Z}_2$ model}
\label{num}

In this section we analyze the potential of the $O(2) \otimes \mathbbm{Z}_2$ model while varying 
two of its parameters.
Starting with 
the parameters \eqref{newparameters} we fix the quadratic parameters 
$\mu_0$, $\mu_{12}$, $\mu_m$ by 
\eqref{tadpole}.
Quantitatively, we choose the quartic parameters motivated by the central point
given in~\cite{Grimus:2008dr} with a variation 
of the two parameters $a_1 \in [0, 5]$ and $a_2\in [-3,3]$
in steps of $0.2$:
\begin{equation} \label{para}
\begin{split}
&a_1 \in [0, 5], \;
a_2 \in [-3,3], \;
a_3=-5, \;
a_4=-0.0474, \\
&
a_5=1.5, \;
a_6=2, \;
a_7=3, \\
&v_1/v_2 = 105.66/1776.82, \;
v_3 = v_0/\sqrt{2},
\end{split}
\end{equation}
with the masses of the muon, $m_\mu$, and tau, $m_\tau$, given in \cite{Agashe:2014kda}.
The central point in particular passes the electroweak precision observables --
for details, see~\cite{Grimus:2008dr}.

\begin{figure*}
\includegraphics[width=0.8\textwidth]{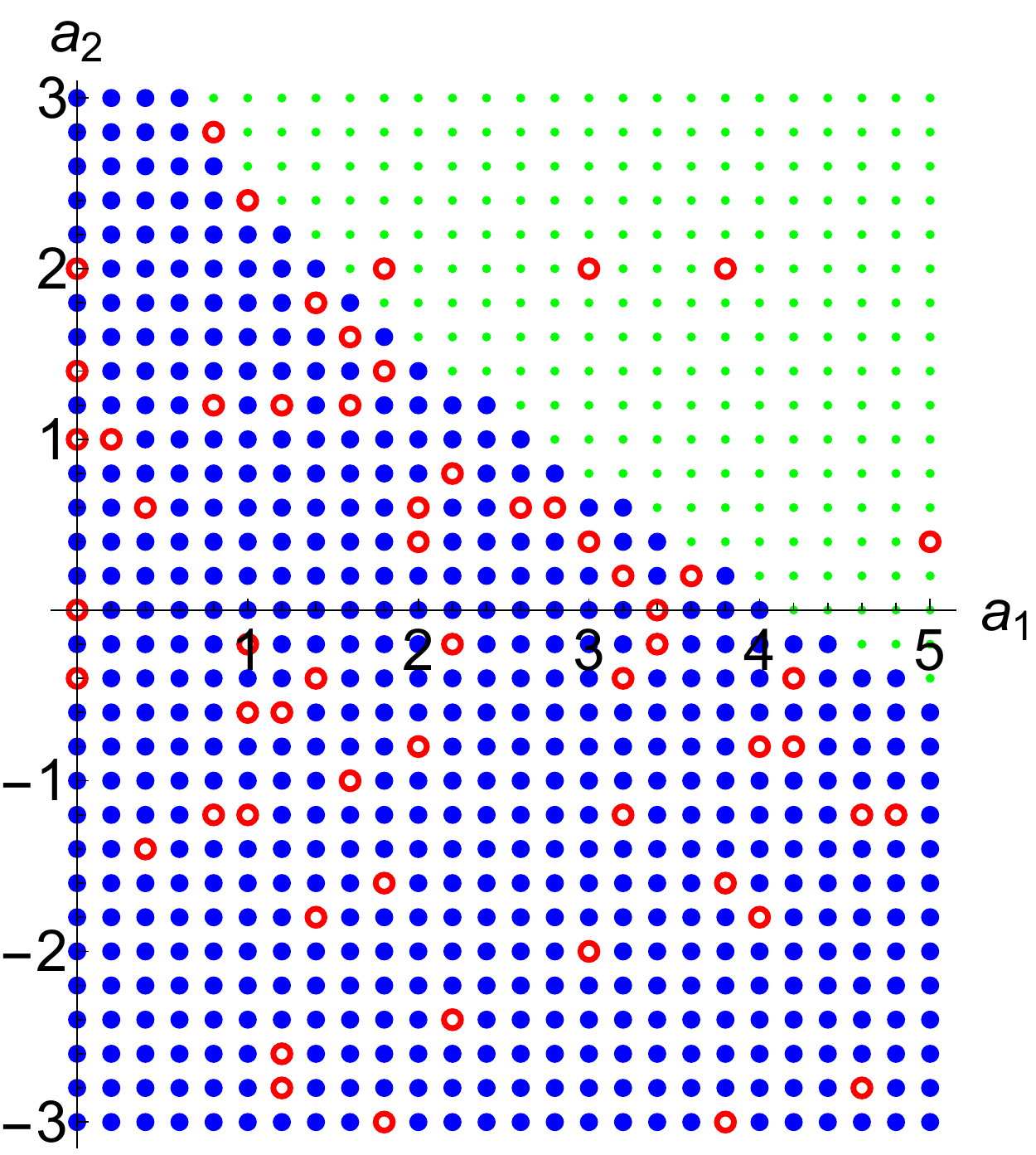}
\caption{\label{a1a2} Stability and stationarity solutions of the
3HDM Higgs potential, varying the two quartic parameters $a_1$ and $a_2$ of
the potential~\eqref{VO2}.
The other quartic parameters are set to $a_3=-5$, $a_4=-0.0474$, $a_5=1.5$, $a_6=2$, $a_7=3$.
The quadratic parameters are fixed by the equations~\eqref{tadpole}, 
the vacuum expectation value $v_0$, and the ratio $v_1/v_2=m_\mu/m_\tau$ as well as $v_3=v_0/\sqrt{2}$.
The larger full circles (blue) show points where the potential is unstable. 
The open circles (red) show parameters
where no correct electroweak symmetry breaking appears or the vacuum expectation value
$v_0$ is unequal to 246 GeV. Finally, the
small dots (green) have a viable global minimum corresponding to the correct
vacuum expectation value.}
\end{figure*}

First, we study the stability of the potential. Therefore, we separate the potential
into the quadratic and quartic parts $V=K_0 J_2 + K_0^2 J_4$, with $J_2$ and $J_4$
given by
\begin{equation} \label{J24}
\begin{split}
J_2(k_a) = &\frac{\mu_0 + 2 \mu_{12}}{\sqrt{6}} 
+ \left( \frac{\mu_{12} - \mu_0}{\sqrt{3}} \right) k_8
+ \mu_m k_1, \\
J_4(k_a) = & \frac{1}{6}(a_1 + 2 a_2 + 2 a_5 + a_6)\\
& + \frac{1}{3 \sqrt{2}} (-2 a_1 - a_2 + 2 a_5 + a_6) k_8
+ \frac{a_7}{4} (k_1^2+k_2^2) \\
&+ \frac{1}{4} (2 a_5 - a_6 ) k_3^2 
+ \frac{a_3}{4} (k_4^2+k_5^2+k_6^2+k_7^2)\\
&+ \frac{\re (a_4)}{2} (k_4 k_6 - k_5 k_7)
+ \frac{\im (a_4)}{2} (k_4 k_7 + k_5 k_6)\\
&+ \frac{1}{12}(4 a_1 - 4 a_2 + 2 a_5 + a_6) k_8^2,
\end{split}
\end{equation}
with the vector components $k_a= K_a/K_0$, $a=1,\ldots,8$ defined for $K_0 \neq 0$.
For $K_0=0$ the potential vanishes.
The stationary points of~$J_4(k_a)$ corresponding to a matrix $\twomat{K}$ with rank~2
are obtained from 
\begin{equation}  \label{stab2}
\begin{split}
& \nabla_{k_1,\ldots,k_8} \bigg[
J_4(k_a)-
u \bigg( 
\det ( \sqrt{2/3} \unitmatrix_3 + k_a \lambda_a)
\bigg)
\bigg] =0,\\ 
& \det ( \sqrt{2/3} \unitmatrix_3 + k_a \lambda_a) = 0, \\
&2- k_a k_a > 0 ,
\end{split}
\end{equation}
and the stationary points corresponding to a 
matrix $\twomat{K}$ with rank~1
are obtained from
\begin{equation} \label{stab1}
\begin{split}
& \nabla_{\tvec{w}^\dagger} \bigg[ J_4(k_a(\tvec{w}^\dagger, \tvec{w})-
u \left(\tvec{w}^\dagger \tvec{w} - 1 \right) \big] = 0,\\
& \tvec{w}^\dagger \tvec{w} - 1 =0
\end{split}
\end{equation}
where we express the vector components $k_a$ by
\begin{equation}
k_a(\tvec{w}^\dagger, \tvec{w}) = \sqrt{\frac{3}{2}} \tvec{w}^\dagger \lambda_\alpha \tvec{w}
\end{equation}
and $\tvec{w}$, $\tvec{w}^\dagger$ are three-component complex vectors. 
If we have for all real solutions $k_a$ of the 
systems of polynomial equations \eqref{stab2} and \eqref{stab1}
$J_4(k_a)>0$, or at least $J_4(k_a)=0$ but then in
addition $J_2(k_a)\ge0$, the potential is stable.
In other words, if there is for a given initial parameter set one solution
with $J_4(k_a)<0$ or 
$J_4(k_a)=0$ but in
addition $J_2(k_a)<0$ the potential is unstable.
The unstable cases for the variation of parameters~\eqref{para}
are denoted by the larger full disks (blue) in Fig.~\ref{a1a2}. For all other
values of parameters the potential is stable.

Let us note that 
the quartic parameters $a_1$ and $a_2$ appear as coefficients of $(\varphi_3^\dagger \varphi_3)^2$
and $\varphi_3^\dagger \varphi_3 \big( \varphi_1^\dagger \varphi_1 + \varphi_2^\dagger \varphi_2)$, respectively,
in the potential~\eqref{VO2}.
Therefore it is evident that the potential
is unstable for small values of $a_1$ and too negative values for $a_2$.\\

Having determined parameter sets giving a stable potential
we proceed with
the study of the stationary points in these cases. 
We systematically look for all stationary points of the potential. 
To this end we have to solve the following
systems of polynomial equations, 
corresponding to solutions which break electroweak symmetry partially (conserving
the elecromagnetic \emgroup symmetry), and solutions which 
break the electroweak symmetry completely.

The stationary solutions with full electroweak symmetry breaking,
corresponding to  
stationarity matrices $\twomat{K}= K_\alpha \lambda_\alpha /2$ of rank~2 are
obtained from
\begin{equation}\label{stat2}
\begin{split}
& \nabla_{K_0,\ldots,K_8} \bigg[ V(K_0,\ldots,K_8) - 
u\; 
\det(\twomat{K})
\bigg] =0, \\  
&2 K_0^2 - K_a K_a >0,\\ 
&\det(\twomat{K})=0,\\
& K_0 >0.
\end{split}
\end{equation}

The stationary solutions with partial electroweak symmetry breaking,
corresponding to stationarity matrices $\twomat{K} = K_\alpha \lambda_\alpha /2$
of rank~1 are obtained from 
\begin{equation} \label{stat1}
\begin{split}
& \nabla_{\tvec{w}^\dagger, K_0} \big[
V( K_\alpha(K_0, \tvec{w}^\dagger, \tvec{w} )) - u ( \tvec{w}^\dagger  \tvec{w} -1 ) 
\big]= 0,\\
& \tvec{w}^\dagger  \tvec{w} -1 = 0,\\
& K_0 >0
\end{split}
\end{equation}
where we express the bilinears $K_\alpha$ in terms of $K_0$ and the three component 
complex vectors $\tvec{w}$
and $\tvec{w}^\dagger$,
\begin{equation} \label{Ktow}
K_\alpha(K_0, \tvec{w}^\dagger, \tvec{w} ) = 
\sqrt{\frac{3}{2}} K_0 \tvec{w}^\dagger \lambda_\alpha \tvec{w},\; \alpha = 0, \ldots, 8.
\end{equation}
In \eqref{stat2} and \eqref{stat1}, $u$ is a Lagrange multiplier, respectively.

In addition there is always a solution for a vanishing potential, corresponding
to an unbroken electroweak symmetry.

The global minimum, that is, the vacuum, 
is given by the stationary point or points with
the deepest potential value. 
In case this solution originates from the set~\eqref{stat1}, that is,
when we do have electroweak symmetry breaking \eweakgroup $\to$ \emgroup,
we can directly calculate the vacuum-expectation value of this minimum, 
$v_0^2 = \sqrt{6} K_0$,
and verify that it coincides with \eqref{vev}.
 
Depending on the variation of the parameters $a_1$ and $a_2$
we  detect the viable global minima. These cases are marked by little (green)
dots in Fig.~\ref{a1a2}.
In case the deepest potential value does not correspond to
the correct electroweak symmetry breaking or 
does not give the correct vacuum expectation value, these parameter
points are denoted by a circle (red) in Fig.~\ref{a1a2}.

As we can see by the scattering of points, we typically find valid
parameters for $a_1$ and $a_2$ not too small. 
However, the pattern of valid points appears very sensitive
to the parameter values. This, of course, is a consequence of the rather
involved potential~\eqref{VO2}.

Eventually, let us remark on the technical aspects to solve the
rather involved systems of equations - on the one hand for the
study of stability~\eqref{stab2}, \eqref{stab1}, 
and on the other hand for the study of stationarity~\eqref{stat2}, \eqref{stat1}.
We apply for all the polynomial systems of equations
the homotopy continuation approach as implemented in the 
PHCpack package~\cite{phcpack}. For a brief introduction to
the homotopy continuation method, see for instance~\cite{Maniatis:2012ex}. 

In the case of the systems of equations \eqref{stab1}, \eqref{stat1},
we decompose the three component complex vectors $\tvec{w}^\dagger$, $\tvec{w}$ into
real and imaginary parts. 
In turn we split every equation in
the sets into
real and imaginary parts. In this way, all indeterminants in all systems of equations
have to be real, and we discard all nonreal solutions.
Practically, we treat any solution as real if the imaginary part of any of the inderminants 
has an amount smaller than~0.001. 

Let us study some details of the potential in case it
has the correct electroweak symmetry breaking
 $\eweakgroup \to \emgroup$. This is conveniently done
in a new basis
\begin{equation}
\varphi_i'(x)= U_{ij} \varphi_j(x), \quad i,j \in \{1,2,3\}
\end{equation}
in which only the field $\varphi_3'(x)$ gets a nonvanishing vacuum expectation value $v_0/\sqrt{2}$.
The unitary matrix $U$ is determined by two rotation angles, which have to fulfill
\begin{equation}
\begin{split}
&v_1 = \sin (\beta_2) \sin (\beta_1)  v_0,\\
&v_2 = \sin (\beta_2) \cos (\beta_1)  v_0,\\
&v_3 = \cos (\beta_2) v_0
\end{split}
\end{equation}
with $\beta_1 \in [0,2 \pi [$, $\beta_2 \in [0,\pi ]$ and we have
in particular $v_1^2+v_2^2+v_3^2=v_0^2$.
In the new basis, the bilinear parameter vector  $(\xi_\alpha)$ from \eqref{paraK} becomes
\begin{small}
\begin{equation} \label{xiprime}
(\xi_\alpha')=
\begin{pmatrix}
\xi_0\\
\cos (2 \beta_1) \cos (\beta_2) \xi_1\\
0\\
\sin(\beta_1)\cos(\beta_1)(1+\cos^2(\beta_2)) \xi_1 + \sqrt{3}/2 \sin^2(\beta_2) \xi_8\\
-\cos (2 \beta_1) \sin (\beta_2) \xi_1\\
0\\
\sin(\beta_2) \cos(\beta_2) (\sin(2 \beta_1) \xi_1 - \sqrt{3} \xi_8)\\
0\\
\sqrt{3} \cos(\beta_1)\sin(\beta_1) \sin^2(\beta_2) \xi_1 + \frac{1}{4}(1+3 \cos(2\beta_2))\xi_8
\end{pmatrix}
\end{equation}
\end{small}
The correct global minimum has, in terms of these parameter components,
the potential value
\begin{equation} \label{V0}
\langle V \rangle = \frac{v_0^2}{2 \sqrt{6}} \left (\xi'_0 - \sqrt{2} \xi'_8 \right).
\end{equation}
This result serves as a good cross-check of the numerical study.
In particular, the potential value at vacuum has to be nonpositive in order
not to lie above the unbroken minimum, which is given by a vanishing potential.

The physical Higgs bosons follow from the diagonalization of the charged and
neutral squared Higgs mass matrices.
The neutral field of the third Higgs boson $\varphi_3$, that is, $h^0(x)$, 
is a mass eigenstate if the parameters fulfill
\begin{equation} 
\xi'_4=\xi'_5=\xi'_6=\xi_7'=0.
\end{equation}
Therefore with a view on \eqref{xiprime}, we find the
explicit conditions for the $O(2) \otimes \mathbbm{Z}_2$ model, where
the neutral component $h^0(x)$ is {\em aligned} with
the vacuum expectation value. 
 The mass squared of the neutral component is in this case 
\begin{equation} \label{HiggsLHC}
m_{h^0}^2 = \frac{4}{\sqrt{3}} \big(\xi'_8 - \frac{1}{\sqrt{2}} \xi'_0  \big).
\end{equation}
{\em Alignment} thus requires that either $v_1=v_2$
 or $\mu_m =0$ which is not the case, since the ratio of the two
 vacuum expectation values $v_1$ and $v_2$ is fixed by the ratio of the muon and tau masses,
 and a nonvanishing parameter $\mu_m$ is required in order to
 break the $O(2) \otimes \mathbbm{Z}_2$ symmetry softly.

Let us comment on the neutrino mixing angles 
$\theta_{13}=0$ and $\theta_{23}= \pi/4$. Since these
values seem not to be exactly fulfilled experimentally, we mention
that deviations may be achieved by imposing further soft-breaking
terms in the potential \eqref{VO2}. 
We have seen that the value and the nature of the global minimum
is rather sensitive to small changes of the potential. Therefore,
we expect that definite results would require
a separate study of the changed potential. However, since additional
soft-breaking terms do not affect the quartic part of the potential,
and, in particular, we have found that for large parts of parameter
space stability follows from the quartic terms alone, we
expect stability also in the respective cases of a potential imposing additional
soft-breaking terms.

We would like to mention that the parameters $a_2$, $a_3$, and $a_4$ couple
the neutral boson $h^0(x)$ with the other two generations of doublets. 
In general this may lead to changed phenomenology in Higgs boson production and 
decay of the $h^0(x)$ field.
For some investigation of this we refer
to \cite{Grimus:2008dr}. A further detailed study for instance,
with respect to a possible enhancement of the $h^0(x)$ decay rate into a pair of photons
is left for future work.

\section{Conclusions}

The $O(2) \otimes \mathbbm{Z}_2$ model~\cite{Grimus:2008dr} introduces three Higgs-boson doublets 
accompanied by an appropriate assignment of the elementary particles to
irreducible representations of the group. In this way a neutrino mass matrix
is generated which corresponds to mixing angles which are close
to the experimental measurements.
However, even though the symmetry restricts the model, the Higgs potential
appears to be rather involved. Nevertheless, the recently
introduced methods to study any three-Higgs doublet model~\cite{Maniatis:2014oza}
were applied to study the potential in detail. We have investigated stability,
the stationary points, and electroweak symmetry breaking
of the Higgs potential by solving the corresponding stationary equations, employing 
polynomial homotopy continuation. The method numerically 
finds all the isolated complex solutions out of which we have extracted the physical real solutions. 
We have scanned over a range of values of the potential parameters.
As expected, for too low
values of the quartic parameters typically an unstable potential is encountered.
For parameter values, corresponding to a stable potential, the global
minimum was detected.
Our study reveals that in this model there are viable parameters
corresponding to a stable global minimum with
correct electroweak symmetry breaking.


\section*{Acknowledgement}
We would like to thank Luis Lavoura and Walter Grimus as well as the
unknown referees very much
for valuable comments.
D. M. was supported by a DARPA Young Faculty Award 
and an Australian Research Council DECRA fellowship.
C. R. and M. M. were supported partly by the 
Chilean research project FONDECYT, Project No.
1140781, and No. 
1140568,respectively. as well as the group  {\em F\'{i}sica
de Altas Energias} of the Universidad del B\'{i}o-B\'{i}o.


\begin{thebibliography}{99}

\bibitem{Agashe:2014kda} 
  K.~A.~Olive {\it et al.}  [Particle Data Group Collaboration],
  ``Review of Particle Physics,''
  Chin.\ Phys.\ C {\bf 38}, 090001 (2014).


\bibitem{GonzalezGarcia:2012sz} 
  M.~C.~Gonzalez-Garcia, M.~Maltoni, J.~Salvado and T.~Schwetz,
  ``Global fit to three neutrino mixing: critical look at present precision,''
  J. High Energy Phys. {\bf 12}, 123 (2012)
  [arXiv:1209.3023 [hep-ph]].


\bibitem{King:2013eh} 
  S.~F.~King and C.~Luhn,
  ``Neutrino Mass and Mixing with Discrete Symmetry,''
  Rept.\ Prog.\ Phys.\  {\bf 76}, 056201 (2013)
  [arXiv:1301.1340 [hep-ph]].


\bibitem{Ma:2005pd} 
  E.~Ma,
  ``Neutrino mass matrix from S(4) symmetry,''
  Phys.\ Lett.\ B {\bf 632}, 352 (2006)
  [arXiv:hep-ph/0508231].


\bibitem{Ma:2009wi} 
  E.~Ma,
  ``Neutrino Tribimaximal Mixing from A(4) Alone,''
  Mod.\ Phys.\ Lett.\ A {\bf 25}, 2215 (2010)
  [arXiv:0908.3165 [hep-ph]].

\bibitem{Felipe:2013vwa} 
  R.~Gonzalez Felipe, H.~Serodio and J.~P.~Silva,
  ``Neutrino masses and mixing in A4 models with three Higgs doublets,''
  Phys.\ Rev.\ D {\bf 88}, 015015 (2013)
  [arXiv:1304.3468 [hep-ph]].


\bibitem{Ma:2013xqa} 
  E.~Ma,
  ``Neutrino Mixing and Geometric CP Violation with $\Delta(27)$ Symmetry,''
  Phys.\ Lett.\ B {\bf 723}, 161 (2013)
  [arXiv:1304.1603 [hep-ph]].

\bibitem{Grimus:2008dr} 
  W.~Grimus, L.~Lavoura, and D.~Neubauer,
  ``A light pseudoscalar in a model with lepton family symmetry O(2),''
  J. High Energy Phys. {\bf 07}, 051 (2008)
  [arXiv:0805.1175 [hep-ph]].


\bibitem{Altarelli:2010gt} 
  G.~Altarelli and F.~Feruglio,
  ``Discrete Flavor Symmetries and Models of Neutrino Mixing,''
  Rev.\ Mod.\ Phys.\  {\bf 82}, 2701 (2010)
  [arXiv:1002.0211 [hep-ph]].


\bibitem{Grimus:2007if} 
  W.~Grimus, L.~Lavoura, O.~M.~Ogreid and P.~Osland,
  ``A precision constraint on multi-Higgs-doublet models,''
  \mbox{J.\ Phys.\ G {\bf 35}, 075001 (2008)}
  \mbox{[arXiv:0711.4022 [hep-ph]].}

\bibitem{Nagel:2004sw}
  F.~Nagel,
  ``New aspects of gauge-boson couplings and the Higgs sector,''
Ph.D.-thesis, Heidelberg University (2004).

\bibitem{Maniatis:2006fs} 
  M.~Maniatis, A.~von Manteuffel, O.~Nachtmann and F.~Nagel,
  ``Stability and symmetry breaking in the general two-Higgs-doublet model,''
  Eur.\ Phys.\ J.\ C {\bf 48}, 805 (2006)
  [arXiv:hep-ph/0605184].

\bibitem{Nishi:2006tg} 
  C.~C.~Nishi,
  ``CP violation conditions in $N$-Higgs-doublet potentials,''
  Phys.\ Rev.\ D {\bf 74}, 036003 (2006)
  [Erratum {\bf 76}, 119901 (2007)]
  [arXiv:hep-ph/0605153].

  
\bibitem{Maniatis:2014oza} 
  M.~Maniatis and O.~Nachtmann,
  ``Stability and symmetry breaking in the general three-Higgs-doublet model,''
  \mbox{J. High Energy Phys. {\bf 02}, 058 (2015)}
  \mbox{[arXiv:1408.6833 [hep-ph]].}
  

\bibitem{phcpack}
  J.~Verschelde, 
 ``Algorithm 795: PHCpack: A General Purpose Solver for Polynomial systems 
 by Homotopy Continuation,''
 ACM Trans. Math. Software, {\bf 25}, 2 (1999).


\bibitem{Maniatis:2012ex} 
  M.~Maniatis and D.~Mehta,
  ``Minimizing Higgs Potentials via Numerical Polynomial Homotopy Continuation,''
  Eur.\ Phys.\ J.\ Plus {\bf 127}, 91 (2012)
  [arXiv:1203.0409 [hep-ph]].


\end{thebibliography}
\end{document}